\documentclass[11pt,reqno]{amsart}
\usepackage{amsmath,amssymb,tabularx,setspace,color,multirow,graphics,graphicx}
\usepackage{amsfonts,amssymb,amsthm}
\usepackage{setspace}
\usepackage[margin=3.5cm]{geometry}
\usepackage[colorlinks=true, citecolor=blue]{hyperref}

\usepackage{adjustbox}
\usepackage{float}
\usepackage{placeins}
\usepackage{booktabs}
\usepackage{longtable}
\usepackage{caption}
\usepackage{subcaption}
\usepackage{multirow}
\usepackage{url}

\newtheorem{definition}{Definition}
\newtheorem{remark}{Remark}

\newtheorem{Definition}{Definition}[section]
\newtheorem{Corollary}{Corollary}[section]
\newtheorem{Remark}{Remark}[section]
\newtheorem{Theorem}{Theorem}[section]

\newtheorem{Example}{Example}[section]

\newtheorem{Pro}{Properties}[section]

\begin{document}
\doublespace
\title[]{On Weighted Entropy Generating Function}
\author[]%
{S\lowercase{mitha} S.$^{\lowercase{a}}$, M\lowercase{ary} A\lowercase{ndrews$^{\lowercase{a, b}}$ and } S\lowercase{udheesh} K. K\lowercase{attumannil}$^{\lowercase{c}}$   \\
 $^{\lowercase{a}}$K E C\lowercase{ollege} M\lowercase{annanam,} K\lowercase{erala}, I\lowercase{ndia},\\
 $^{\lowercase{b}}$S H C\lowercase{ollege} T\lowercase{hevara,} K\lowercase{erala}, I\lowercase{ndia},\\$^{\lowercase{c}}$I\lowercase{ndian} S\lowercase{tatistical} I\lowercase{nstitute},
  C\lowercase{hennai}, I\lowercase{ndia.}}
\maketitle
%
\begin{abstract}
 In this paper, we study the properties of the weighted entropy generating function (WEGF). We also introduce the weighted residual entropy generating function (WREGF) and establish some characterization results based on its connections with the hazard rate and the mean residual life function. Furthermore, we propose two new classes of life distributions derived from WREGF. We also study the non-parametric estimation of WREGF. A non-parametric test for the Pareto type I distribution is developed based on entropy characterization.  To evaluate the performance of the test statistics, we conduct an extensive Monte Carlo simulation study. Finally, we apply the proposed method to two real-life datasets.
\end{abstract}

\section{Introduction}
 The notion of entropy introduced by Shannon (1948) plays a vital role not only in the area of information theory but also in various other scientific disciplines. When considering a non-negative random variable $X$ with a probability density function $f(x)$, Shannon's entropy is defined as
\begin{eqnarray}\label{entropy}
H(X)&=&-\int _0^\infty f(x) \log f(x)~dx,
\end{eqnarray}
where $\log$ denotes the natural logarithm. It measures the expected level of uncertainty within a probability density function concerning the predictability of an outcome related to variable $X$. For more properties and applications of (\ref{entropy}), see Shannon (1948) and Wiener (1948).
\par The study of entropy based measures in the context of lifetime distributions has evolved significantly, with residual entropy emerging as a key concept in reliability theory and information analysis. Shannon entropy is not useful for measuring
the uncertainty about the remaining lifetime of a system.  Building on this, Ebrahimi and Pellery (1995) was among the first to formally address uncertainty in residual lifetime distributions, proposing residual entropy as a tool to measure the uncertainty associated with the remaining lifetime of a component or system as
\begin{eqnarray}\label{Resentropy}
H(X;t)&=&-\int _t^\infty \frac{f(x)}{\bar{F}(t)} \log \frac{f(x)}{\bar{F}(t)}~dx.
\end{eqnarray}


\par It is a well-established fact that if the moments exist, the successive derivatives of the moment generating function at point 0 provide the successive moments of a probability distribution. In a correspondence, Golomb (1966) introduced the information generating function of a probability distribution, referred to as the entropy generating function, and defined it as follows
\begin{eqnarray}\label{egf}
B(s)&=& \int_0 ^\infty f^s(x) dx,~s> 0, s\neq 1.\label{egf}
\end{eqnarray}
Smitha et al. (2024) proposed a dynamic version of the generating function given in (\ref{egf}) as
\begin{equation}\label{degf}
B_t(s)=\int _t ^\infty {\left( \frac{f(x)}{\bar{F}(t)}\right) }^s dx,~s> 0, s\neq 1.
\end{equation}
\par The first derivative of $B_t(s)$ at $s=1$ and with negative sign gives the residual entropy function introduced by Ebrahimi and Pellerey (1995). They also characterized certain lifetime distributions using the functional form of this measure and proposed a non-parametric estimator for the new measure.
\par In many real-life situations, standard probability distributions may not fit the data well. To deal with this, weighted distributions are used. In recent years, this concept has been widely applied in several areas of statistics, including the analysis of family size, human heredity, global population studies, renewal theory, biomedical research, statistical ecology, and reliability modeling. Shannon entropy gives equal weight to the occurrence of every event. Belis and Guiasu (1968) proposed weighted entropy, which is a generalization of classical entropy and is defined as
\begin{eqnarray}\label{wentropy}
H^w(X)&=& -\int _0^\infty x f(x) \log f(x)~dx.
\end{eqnarray}
\par Di Crescenzo and Longobardi (2006) introduced the notions of weighted residual entropy and weighted past entropy that are suitable to describe dynamic information of random lifetimes. They pointed out the application of these measures in reliability, neurobiology etc. Misagh and Yari (2011) studied the weighted differential information measure for two sided truncated random variables. Building on the concept of cumulative residual entropy (CRE), Mirali et al. (2017) introduced a new information measure known as weighted cumulative residual entropy (WCRE).
\par Motivated by the usefulness of weighted information measures and generating functions, this paper presents an extensive study on the properties of the weighted entropy generating function (WEGF) proposed by Saha and Kayal (2024) and introduces the weighted residual entropy generating function (WREGF). A detailed study based on the WREGF is also integrated.
\par This paper is organized as follows. The properties of the weighted entropy generating function have been studied in Section 2. In Section 3, we propose the weighted residual entropy generating function (WREGF). The properties including linear transformation, bounds and WREGF of some well known lifetime distributions are also discussed. In Section 4, we give some characterization results based on WREGF. Section 5 introduces two new classes of life distribution based on WREGF. A new goodness of fit test for Pareto distribution is detailed in Section 6. Section 7 explains the concept of linear combination of kernels and also discusses the consistency of the estimator of the test statistic. Monte Carlo simulation studies are carried out in Section 8 using R software. Section 9 presents data analysis using two datasets to demonstrate the practical applicability of the proposed test statistic.  Conclusions are given in Section 10.
\section {Weighted Entropy Generating Function  and its properties}
\par Golomb’s entropy generating function (Golomb (1966)) treats all the outcomes of a probabilistic experiment as equally important, irrespective of their significance to the goal of the experimenter. However, in practice, certain events may carry more importance than others. We noted that  Saha and Kayal (2024) proposed the General Weighted Information Generating Function (GWIGF) for continuous probability distributions mentioned in (\ref{wegf}) .
In this section, we  extensively studied the properties of the weighted entropy generating function, which considers both the objective probabilities and the real qualitative weights of the elementary events.
\begin{Definition}
  Let $X$ be an absolutely continuous non-negative random variable representing the lifetime of a system with probability density function $f(x)$, distribution function $F(x)$ and survival function $\bar{F}(x)$, then the weighted entropy generating function $B_s(W,X)$ associated with the random variable $X$ is defined as
\begin{eqnarray}\nonumber
B_s(W,X)&=& \int_0 ^\infty x f^s(x) dx,~s\geq 0, s\neq 1.\\
&=& E(X f^{s-1}(X)).\label{wegf}
\end{eqnarray}
\end{Definition}
The factor $X$ in the integrand of (\ref{wegf}) represents a weight which linearly emphasizes the occurrence of the event ${X=x}$. This is a ``length-biased” shift-dependent information measure assigning greater importance to larger values of $X$. When the weight function depends on the length of the component, the resulting distribution is called length-biased weighted function (Misagh, 2016).  Differentiating (\ref{wegf}) with respect to $s$, and putting $s=1$ ,we get
 \begin{eqnarray}\nonumber
{B_s}^{'}(W,X)&=& \int_0 ^\infty x f(x)\log f(x) dx.
\end{eqnarray}
This is the negative of weighted entropy introduced by Belis and Guiasu (1968) given in (\ref{wentropy}).

Next, we evaluate the WEGF of some random variables. It should be noted that there exists a relationship between the weighted entropy generating function and the entropy generating function in terms of the arithmetic mean.
\begin{Example}
If $X$ is exponentially distributed with parameter $\lambda>0$ , then
$$
B_{s}(W,X)=\frac{\lambda^{s-2}}{s^{2}}.
$$
 That is $$ s.B_{s}(W,X)=E(X).B_{s}(X).$$
\end{Example}
\begin{Example}\label{ex2.2}
If $X$ is uniformly distributed on $[a,b]$, then
\begin{eqnarray}
 B_{s}(W,X)=\frac{(b+a)}{2(b-a)^{s-1}}=E(X).B_{s}(X).
\end{eqnarray}
\end{Example}
\begin{Remark}
 In general $B_{s}(W,X)$ can be larger or smaller than $B_{s}(X)$. For example, if X follows  Uniform $[a,b]$ , from (\ref{wegf}), it follows
\[ B_{s}(W,X) \geq B_{s}(X), \text{ when } \mathbb{E}(X) > 1 \]
\[ B_{s}(W,X) \leq B_{s}(X), \text{ when } \mathbb{E}(X) < 1 .\]
\end{Remark}
Table ~\ref{h1} gives the expression of $B_s(W,X)$ for some well-known distributions.

\begin{table}[ht]
\centering
\caption{$B_s(W, X)$ for some well-known distributions.}
\label{h1}
\begin{tabular}{@{}p{4.8cm}p{4.5cm}p{5.5cm}@{}}
\toprule
\textbf{Distribution} & \textbf{Density Function} & \textbf{WEGF} \\
\midrule
Lomax distribution $(m)$ & $\displaystyle\frac{m}{(1 + x)^{1 + m}},\; m > 0, x \geq 0$ & $\displaystyle\frac{m^s}{[(1 + m)s - 1][(1 + m)s - 2]}$ \\
\addlinespace[0.75em]
Power distribution $(c)$ & $c x^{c - 1},\; c>0, 0 < x < 1$ & $\displaystyle\frac{c^s}{(c - 1)s + 2}$ \\
\addlinespace[0.75em]
Pareto distribution $(\alpha)$ & $\alpha x^{-(\alpha + 1)},\; \alpha > 0, x \geq 1$ & $\displaystyle\frac{\alpha^s}{(\alpha + 1)s-2 }$ \\
\addlinespace[0.75em]
Uniform distribution $(a, b)$ & $\displaystyle\frac{1}{b - a},\; a < x < b$ & $\displaystyle\frac{b^2 - a^2}{2(b - a)^s}$ \\
\addlinespace[0.75em]
Exponential distribution $(\lambda)$ & $\lambda e^{-\lambda x},\; \lambda > 0, x \geq 0$ & $\displaystyle\frac{1}{s^2 \lambda^{2 - s}}$ \\
\bottomrule
\end{tabular}
\end{table}

\FloatBarrier
\begin{Remark}
It is to be noted that different distributions may have identical WEGF. For instance, if $s=2$ and $X$ is uniformly distributed over [0,1] and $Y$ follows Pareto distribution with parameter $\alpha =1$, then from Table 1, one can obtain
\begin{align*}
    B_s(W,X)=B_s(W,Y)=\frac{1}{2}.
\end{align*}
\end{Remark}
\begin{Pro}
 For the random variable $X$, let $Y=aX+b$ with $a>0$, $b>0$ we get
\[ B_s(W,Y)= a^{1-s}(aB_s(W,X)+bB_s(X)).\]
\end{Pro}
\begin{Pro}
If $X$ and $Y$ are independent, then
\begin{align*}
    B_s(W,X,Y)=B_s(W,X)B_s(W,Y).
\end{align*}
\end{Pro}
Next property gives the lower bound of WEGF in terms of the measure of entropy introduced by Shannon (1948).

\begin{Pro}
 Let $X$ be a random variable having Shannon entropy $H(X)$, then WEGF
\[B_s(W,X)\geq e^{((1-s)H(X)+E(\log(X))}.\]
\end{Pro}
The proof of the above properties are omitted for the sake of simplicity.
\section{Weighted residual entropy generating function}

In various fields such as reliability, survival analysis, economics, business, etc., the duration of a study period is a significant variable. In these instances, information generating functions become dynamic, incorporating a time dependency. Inspired by the dynamic nature associated with time-dependent functions, this section introduces the dynamic version of WEGF, which is known as the weighted residual entropy generating function (WREGF). The WREGF is defined as
\begin{eqnarray}\label{WREGF}
B_{s}(W,X; t)=\int_{t}^{\infty}x\left(\frac{f(x)}{\bar{F}(t)}\right)^s dx,~s\geq 0, s\neq 1.
\end{eqnarray}
\begin{Remark}

 Clearly $ B_{s}(W,X; 0)= B_{s}(W,X)$, which is WEGF given in (\ref{wegf}).
\end{Remark}
Next, we evaluate the WREGF of some lifetime distributions.

\begin{table}[ht]
\centering
\caption{$B_s(W, X; t)$ for some well-known distributions.}
\label{tbl4}
\begin{tabular}{@{}p{4.5cm}p{4.5cm}p{5.5cm}@{}}
\toprule
\textbf{Distribution} & \textbf{Density Function } & \textbf{WREGF} \\
\midrule
Uniform distribution $[a, b]$ & $\displaystyle\frac{1}{b-a},\; a \leq x \leq b$ & $\displaystyle\frac{b + t}{2(b - t)^{s - 1}}$ \\
\addlinespace[0.75em]
Exponential distribution $(\lambda)$ & $\lambda e^{-\lambda x},\; x > 0,\; \lambda > 0$ & $\displaystyle\frac{\lambda^s (1 + \lambda s t)}{(\lambda s)^2}$ \\
\addlinespace[0.75em]
Power distribution $(c)$ & $c x^{c - 1},\; 0 < x < 1, c>0$ & $\displaystyle\frac{c^s (1 - t^{(c - 1)s + 2})}{(1 - t^c)^s ((c - 1)s + 2)}$ \\
\addlinespace[0.75em]
Pareto distribution $(\alpha)$ & $\alpha x^{-(\alpha + 1)},\; x \geq 1, \alpha>0$ & $\displaystyle\frac{\alpha^s (1 - t^{-(\alpha + 1)s + 2})}{(\alpha + 1)s-2}$ \\
\addlinespace[0.75em]
Lomax distribution $(m)$ & $\displaystyle\frac{m^s}{(1 + x)^{1 + m}},\; x \geq 0,m>0$ & $\displaystyle\frac{m^s (1 + t)(3t - t(1 + m)s + 1)}{(1 - (1 + m)s)(2 - (1 + m)s)}$ \\
\bottomrule
\end{tabular}
\end{table}

\begin{Pro}
$B_s(W,X)$ holds the relation \[ B_s(W,X)=\int_{0}^{t}\ x~{f^s(x)}~dx+(\bar{F}(t))^sB_s(W,X;t).\]
  \textbf {Proof}:
The result is clear from (\ref{wegf}) and (\ref{WREGF}).
\end{Pro}
In the following property, we analyze the effect of linear transformation on WREGF.
\begin{Pro}
 Consider the random variable, $Y=aX+b$ with $a>0$ and $b>0$, the WREGF of $Y$ can be expressed as
\[ B_s\left(W,Y;\frac{t-b}{a}\right)=\left(a^{1-s}\left(a~B_s\left(W,X;\frac{t-b}{a}\right)\right)+bB_s\left(X;\frac{t-b}{a}\right)\right).\]
\end{Pro}
 \textbf {Proof}:
We have
\[B_s(W,Y;t)=\int_{t}^{\infty}\ y\left(\frac{{f}(y)}{P(Y>t)}\right)^s dy.\]
Here $Y=aX+b$,
so
\[ B_s(W, Y; \tfrac{t - b}{a}) = \int_{\frac{t - b}{a}}^{\infty} (a x + b) \left( \frac{\frac{1}{a} f\left( \frac{x - b}{a} \right)}{\bar{F}\left( \frac{t - b}{a} \right)} \right)^s dx.\]
On simplification, we get the required result.\\
Next, we obtain the relationship between hazard rate and WREGF. By differentiating $B_{s}(W,X;t)$ with respect to $t$, we obtain the following result.
\begin{equation}\label{rent}
B'_{s}(W,X;t)- s h(t) B_{s}(W,X;t)=-t(h(t))^s.
\end{equation}
Next theorem shows the relationship between weighted residual entropy generating function and residual entropy generating function.
\begin{Theorem}

We can also express WREGF as
\[ B_s(W,X;t)= tB_s(X;t)+ \int_{t}^{\infty}\left(\frac{\bar{F}(x)}{\bar{F}(t)}\right)^s B_s(X,y) \ dy, \text{ for all } t>0.\]

 \textbf {Proof}: Consider
\begin{eqnarray}
    \int_{t}^{\infty} x \left(\frac{{f}(x)}{\bar{F}(t)}\right)^s dx   & =& \int_{t}^{\infty}\left (\int_{0}^{x} dy\right)\left(\frac{{f}(x)}{\bar{F}(t)}\right)^sdx\nonumber
    \\
& =&  t \int_{t}^{\infty}\left  (\frac{{f}(x)}{\bar{F}(t)}\right)^s dx+\frac{1}{(\bar{F}(t))^s}\int_{y=t}^{\infty}\left(\int_{x=y}^{\infty}({{f}(x)} ^sdx\right)dy.
\end{eqnarray}Hence, we have
\[ B_s(W,X;t)=t B_s(X;t)+\int_{t}^{\infty} \left (\frac {\bar{F}(y)}{\bar{F}(t)}\right) ^s B_s(X,y) \ dy.\]
The above expression can be written as
\[ B_s(W,X;t)= tB_s(X;t)+ \int_{t}^{\infty}\left(\frac{\bar{F}(x)}{\bar{F}(t)}\right)^s B_s(X,y)\ dy.\]
Hence the theorem.
\end{Theorem}
\par Next theorem shows that under certain conditions $B_s(W,F;t)$ uniquely determines the distribution function.
\begin{Theorem}\label{t1.1}
\textit{Let $F(x)$ be an absolutely continuous distribution function and assume that $B_s(W,F;t)$ is increasing in $t$. Then $B_s(W,F;t)$ uniquely determines $F(t)$.}
\end{Theorem}
{\bf Proof:}
Suppose that $F(x)$ and $G(x)$ are distribution functions such that
\begin{equation}\label{6}
    B_s(W,F;t)=B_s(W,G;t), \,\,\forall t \geq 0
\end{equation}
 Using (\ref{WREGF}), (\ref{6}) becomes
\begin{eqnarray}\label{66}
\frac{1}{({{\bar{F}(t)})}^s}\int_t^\infty x f^s(x) dx&=&\frac{1}{({{\bar{G}(t)})}^s}\int_t^\infty x g^s(x) dx,\label{7}
\end{eqnarray}
where $f(x)$ and $g(x)$ are the probability density functions corresponding to $F(x)$ and $G(x)$ respectively. Now differentiating (\ref{7}) with respect to $t$, we get
\begin{equation}\label{8}
-t~h_1^s(t)+s~h_1(t)~B_s(W,F;t)=-t~h_2^s(t)+s~h_2(t)~B_s(W,G;t).
\end{equation}
To prove $\bar{F}(t)=\bar{G}(t)$, it is enough to prove that $h_1(t)=h_2(t), ~\text{for all}~ t \geq 0$.\\  From (\ref{8}), we get
\begin{eqnarray}
h_1(t)\left[ t~h_1^{s-1}(t)-s~B_t(W,F;t)\right] &=& h_2(t)\left[t~ h_2^{s-1}(t)-s~B_t(W,G;t)\right ]. \label{9}\\\nonumber
\text{From the above equation, we get}\\
\frac{h_1(t)}{h_2(t)}&=&\frac{t~h_2^{s-1}(t)-s~B_s(W,G;t)}{t~h_1^{s-1}(t)-s~B_s(W,F;t)}. \label{90}
\end{eqnarray}
Suppose $h_1(t) > h_2(t)$ with $h_i(t)\neq 0, ~i=$1,2,
so $\frac{h_1(t)}{h_2(t)} > 1.$\\
Then (\ref{90}) becomes
\begin{eqnarray*}
&&t~h_2^{s-1}(t)-s~B_s(W,G;t)>t~h_1^{s-1}(t)-s~B_s(W,F;t).
\end{eqnarray*}Using (\ref{6}), we get $h_2(t)>h_1(t)$, which is a contradiction. \\Similarly, we can show that the inequality  $h_1(t)<h_2(t)$ also leads to the contradiction and gives $h_1(t)=h_2(t)$. Hence, the proof of the theorem.

\section{Characterization results}
In this section, we look into the problem of characterization of probability distributions using the functional form of $B_s(W,F;t)$.
The next theorem characterizes the Weibull distribution in terms of $B_s(W,F;t)$.
\begin{Theorem}\label{t2.1}
\textit{For a non-negative random variable X with density function $f(x)$ having WREGF, $B_s(W,F;t)$,  is independent of $t$, for any $s\neq2$, if and only if $X$ follows Weibull distribution. }
\end{Theorem}
 \textbf {Proof}:
The necessary condition of the theorem can be obtained easily.\\
Let
\[B_s(W,F;t)= A(t).\]
Differentiating with respect to $t$, we get
\begin{equation*}
B_s^{'}(W,F;t)=A^{'}(t).
\end{equation*}
Using the relationship with the hazard rate
\begin{equation*}
s~h(t)B_s(W,F;t)- t~(h(t))^{s}= A^{'}(t).
\end{equation*}
Assume that $A(t)=k,$ a constant, independent of $t$.\\
Using this in the above expression, we get
\[ h(t)=\left(\frac{sk}{t}\right)^{\frac{1}{s-1}}.\]
Thus, the failure rate is proportional to $t^{\frac{1}{1-s}}$, which implies that the random variable $X$ follows the Weibull distribution.
Hence the proof.

The next theorem shows that for $s=2$, constant WREGF characterizes the Pareto distribution.
\begin{Theorem}\label{pareto}
\textit{Assume that $s=2$,  WREGF is independent of $t$ if and only if $X$ follows Pareto distribution with density function}
\begin{equation} \label{16}
f(x)=c~x^{-(c+1)}; x\geq 1.
\end{equation}
\end{Theorem}
\textbf {Proof}:
By direct calculation using (\ref{WREGF}) and (\ref{16})
\[
B_s(W,F;t)={\frac{c^{s}(t^{2-s})}{(c+1)s-2}}
.\]
When $s=2$, we get
\[ B_2(W,F;t)={\frac{c^{2}}{2c}}.\]
Conversely, assume that
\[ B_2(W,F;t)= k.\]
Differentiating with respect to $t$ and using the relationship with hazard rate, we get
\[ 2kh(t)-t~(h(t))^{2}=0.\]
It is clear that $h(t)$ is proportional to $t^{-1}$. This is the characterization of the Pareto distribution.

\section{New Class of Lifetime Distributions}
\doublespacing
In this section, we introduce two new classes of life distributions based on $B_s(W,F;t)$.
\begin{Definition}

A random variable $X$ is said to be smaller than the other random variable $Y$ in WREGF, denoted by $X~{\leq}{WREGF} ~(Y)$ if\\
\[
B_s(W,X;t) \leq B_s(W,Y;t), \quad \text{for all } t > 0.
\]

\end{Definition}
\begin{Definition}
The survival function $\bar{F}$ has increasing (decreasing) weighted entropy generating function for residual life IWREGF (DWREGF) if $B_s(W,F;t)$ increasing (decreasing) in $t$, $t>0$ .\\
This implies that \\
\[ B_s^{'}(W,F;t) \geq(\leq)~ 0.\]
\end{Definition}

In the following theorem, we give some upper (lower) bound for $h(t)$ based on the monotonicity of WREGF.
\begin{Theorem} \label{theorem5}
  \textit{Let the random variable $X$ has IWREGF(DWREGF) then $B_s(W,X;t)$ obtains a lower(Upper) bound as
 \[ B_s(W,X;t)\geq(\leq) {\frac{t}{s}}~ (h(t))^{s-1} .\]}
\end{Theorem}
\textbf {Proof}:
From (\ref{WREGF}) and Definition 5.2, the result is obvious, hence the proof is omitted.

\begin{Example}

Consider exponential distribution with survival function
\[ \bar{F}(x)= e^{-\lambda x}; \lambda>0, X>0 . \]
From Table \ref{tbl4},
\begin{equation} \label{17}
B_s(W,X;t)=\frac{\lambda^s(1+\lambda s t)}{(\lambda~ s)^2
}.\\
\end{equation}
Differentiating (\ref{17}) with respect to $t$ on both sides, we get
\[ B_s^{'}(W,X;t)=\frac{\lambda^{s-1}}{s}.\]
It is clear that $B_s^{'}(W,X;t)>0$ so exponential distribution belongs to the class  IWREGF.
\end{Example}

Next, we obtain the bounds of $B_s(W,X;t)$ in terms of the mean residual life function.
\begin{Theorem}
    \textit{If the random variable $X$ has IWREGF(DWREGF), then the lower(upper) bound as}
 \begin{equation}
B_s(W,X;t)\geq(\leq)\frac{t}{s}\left(\frac{1+m^{'}(t)}{m(t)}\right)^{s-1}.
\end{equation}
\textbf{Proof:}
We have the relationship between the hazard rate and the mean residual life function(mrlf)
 \begin{equation}   \label{19}
h(t)=\frac{1+m^{'}(t)}{m(t)}.
 \end{equation}
Using (\ref{19}) and Theorem \ref{theorem5}, we can have the bounds of the proposed measure in terms of mrlf.
\end{Theorem}

\section{Test for Pareto distribution}
In this section,  we develop a goodness of fit test for Pareto distribution using the characterization result given in Theorem \ref{pareto}. Let $X_1,X_2,\ldots,X_n$ be a random sample of size $n$ from $F$.    Based on this sample we are interested in testing the
null hypothesis
\begin{align*}
    H_0 &: \mathcal{F} \in P(\alpha)
\end{align*}
against the alternative
\begin{align*}
    H_1 &: \mathcal{F} \neq P(\alpha).
\end{align*}where $P(\alpha)$ denotes the Pareto distribution with parameter $\alpha$.
Since $B_2(W,F;t)$ is constant, its first derivative is zero. We use this fact to develop the test.
We consider the departure measure $\Delta(F)$ defined by
$$\Delta(F)=\int_1^{\infty}3xF(x)f^2(x)dx-\int_1^{\infty}xf^2(x)dx. $$ In view of Theorem \ref{pareto}, $\Delta(F)$ is zero under $H_0$ and positive under $H_1$.  We consider test statistics as the plug-in estimator of $\widehat \Delta(F)$. Let $X_{(i)}$ be the order statistics. Hence, the test statistics is given by
\begin{eqnarray*}
    \widehat \Delta&=&\frac{3}{n^2}\sum_{i=1}^{n}iX_{(i)}\widehat f(X_{(i)})-\frac{1}{n}\sum_{i=1}^{n}X_{(i)}\widehat f(X_{(i)})\\
    &=&\frac{1}{n^2}\sum_{i=1}^{n}(3i-n)X_{(i)}\widehat f(X_{(i)}),
\end{eqnarray*}
where $\widehat f(x)$ is the kernel density estimator of $f(x).$ \\
Next, we study the consistency of the proposed test statistic.

Let \( X_1, X_2, \ldots, X_n \) be a random sample. The kernel density estimator for the probability density function \( f(x) \) is given by

\[
\hat{f}(x) = \frac{1}{nh} \sum_{j=1}^{n} k\left( \frac{x - X_j}{h} \right)
.\]
where \( k(\cdot) \) is the kernel function and \( h \) is the bandwidth. The kernel density estimator of \( f(x) \) was introduced by Parzen (1962).

According to Mugdadi and Sani (2020), the kernel function \( k(\cdot) \) in the kernel density estimator can be expressed as a linear combination of kernels,

\[
k(u) = \sum_{i=1}^{p} a_i k_i(u),
\]

where \( p \geq 2 \), \( a_i > 0 \), and each \( k_i \) is a kernel function that is symmetric about zero and satisfies \( \int k_i(u) \, du = 1 \).
Substituting this into the kernel density estimator, we get
\[
\hat{f}(x) = \frac{1}{nh} \sum_{j=1}^{n} \sum_{i=1}^{p} a_i k_i\left( \frac{x - X_j}{h} \right)
.\]

Next, a plug-in estimator is considered. Altman and Léger (1995) estimated the optimal bandwidth using an estimator for $D_2(F)$ as suggested by Hall and Marron (1987). This estimator for

\begin{equation}
D_2(F) = \int f^2(x) W(x) dx
\end{equation}
is given as
\begin{equation} \label{21}
\widehat{D_2}(F) = \frac{1}{n(n-1)} \sum_{i \neq j} \frac{1}{h} k \left(\frac{X_i - X_j}{h} \right) W(X_i).
\end{equation}

Considering the proposed test statistic
\begin{align}
\widehat{\Delta} &= \frac{1}{n^2} \sum_{i=1}^{n} (3i - n) X_{(i)} \, \widehat{f}(X_{(i)}) \nonumber \\
&= \frac{1}{n^3} \sum_{i=1}^{n} (3i - n) X_{(i)} \cdot \left( \frac{1}{h} \sum_{j=1}^{n} k\left( \frac{X_{(i)} - X_j}{h} \right) \right).
\end{align}
The expression in (\ref{21})  resembles the proposed test statistic. It is clear that the proposed test statistic is also a linear combination of kernels. \\
To examine the consistency of the proposed test statistic we will consider a standardized linear combination as mentioned in Ahmad and Ran (2004) , which is given by
 \[
\hat{f}(x;h) = \sum_{j=1}^{p} c_j \hat{f}_j(x;h).
\]

where
\[
 \quad 0 \leq c_j \leq 1 \quad \text{for } j = 1, 2, \ldots, p.
\]
The kernel density estimator $\hat{f}(x; h)$ has the expected value and variance as $n \to \infty$, $h \to 0$, and $nh \to \infty$.

\begin{equation}
E(\hat{f}) = f(x) + \frac{h^2}{2} f''(x) \sum_{j=1}^{p} c_j \sigma_j^2 + o(h^2),
\end{equation}

\begin{equation}
\text{Var}((\hat{f})) = \frac{1}{nh} f(x) R(\psi) + o\left(\frac{1}{nh}\right).
\end{equation}

Since the selected bandwidth $\hat{h}$ satisfies $\hat{h} \xrightarrow{p} 0$ and $n \hat{h} \xrightarrow{p} \infty$ as $n \to \infty$, let $\tilde{f}(x) = \text{lin}(\hat{f}(x; \hat{h}))$ and $\hat{h} = O_p(h)$ as $n \to \infty$. Then, we have,

\begin{equation}
\text{Bias}(\tilde{f}, f) = \frac{h^2}{2} f''(x) \sum_{j=1}^{p} c_j \sigma_j^2 + o(h^2) \to 0 \quad \text{as } n \to \infty,
\end{equation}

\begin{equation}
\text{Var}(\tilde{f}) = \frac{1}{nh} f(x) R(\psi) + o\left(\frac{1}{nh}\right) \to 0 \quad \text{as } n \to \infty.
\end{equation}

This means that the kernel density estimator $\tilde{f}(x)$ from the kernel method is a consistent estimator of $f(x)$.









It is straightforward to see that, to ensure the kernel density estimate is consistent with the underlying density, two conditions on the bandwidth are needed, as $n \to \infty$: $h \to 0$ and $nh \to \infty$. When these two conditions hold, $MSE(\hat{f}(x)) \to 0$, ensuring consistency. On a similar line, we can prove that the variance of our test statistic converges to zero.\\
\par
\begin{figure}[h]
    \centering
    \includegraphics[width=0.8\linewidth]{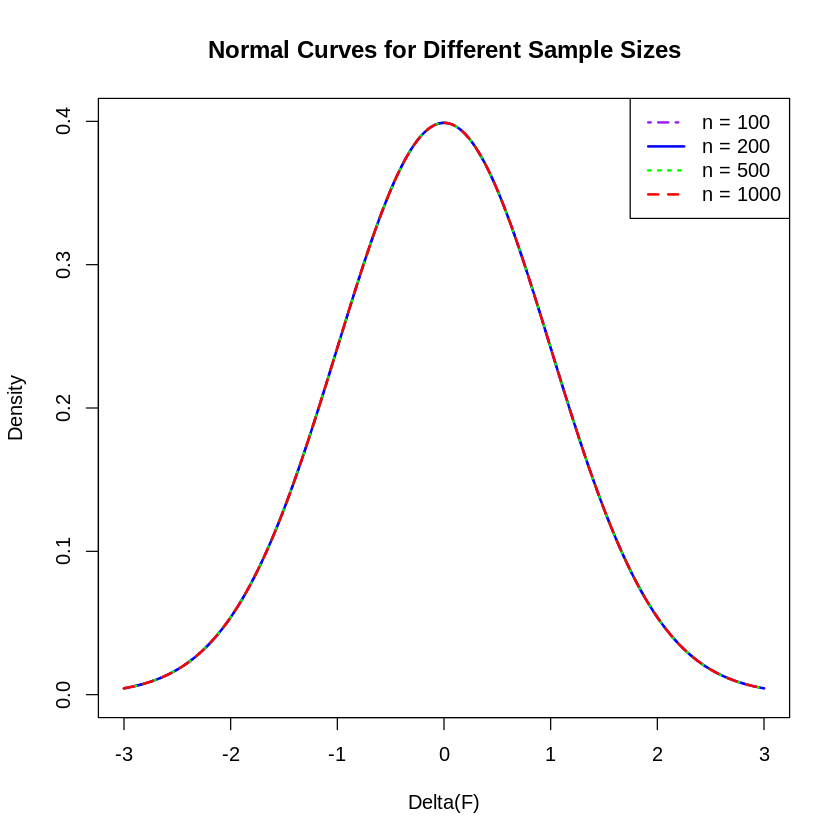}
    \caption{Normal density plot}
    \label{fig:enter-label}
\end{figure}

 Next,  we investigate the asymptotic distribution of the proposed test statistic  using a simulation study,  generated with 10,000 samples of sizes $n=100,200,500,1000$.
 From, Figure ~\ref{fig:enter-label}, it is evident that the limiting distribution of the standardized value of the estimator is standard normal.

\section{Simulation }
Monte Carlo simulations are utilized in this section to evaluate and compare the finite-sample performance of the newly proposed test. We generate lifetimes of different sample sizes, $n =
 10,25,50,75,100$ to calculate the empirical size.
We begin by assuming that the data are generated from a Pareto Type‑I distribution with an unknown parameter $\alpha$, a moment-based estimator
\[
\hat{\alpha} = \frac{\bar{x}}{\bar{x} - 1} .
\]
 To obtain the critical region under $H_0$, a parameteric bootstrap procedure is employed by generating 500 resampled datasets. In each bootstrap replicate, a sample is generated from the Pareto null (using the estimated $\hat{\alpha}$), and the test statistic is computed. The empirical quantiles ($\gamma/2$ and $1-\gamma/2$) of these bootstrap test statistic values define the critical region for a significance level $\gamma$ as $0.5$.  The power of the proposed test is evaluated through Monte Carlo simulation. This is done by generating numerous samples under the alternative hypothesis using various alternative distributions, such as the Gamma, Beta-Exponential, Inverse Beta, Tilted Pareto, Benini, Weibull, Lognormal, and Half-Normal distributions. The experiment is repeated 10,000 times for different sample sizes
$n=10,20,30,40,50$. For each generated sample, the test statistic is computed and compared with the bootstrap critical region; the proportion of times the null hypothesis is rejected (i.e., the test statistic falls outside the critical region) provides an estimate of the power of the test. We also considered some other tests for comparison. Computations and simulations were performed exclusively using R software.
\\

\textbf{1. Zhang's Test Statistics: Zhang (2002) proposed two tests with test
 statistics given by
 }
\begin{enumerate}
  \item
    \[
    Z_{An} = -\frac{1}{n} \sum_{j=1}^{n} \ln\Bigl(1 - F\bigl(X_{(j)}\bigr)\Bigr),
    \]

  \item
    \[
    Z_{Bn} = \sum_{j=1}^{n} \left\{ \frac{-\ln\Bigl(1 - F\bigl(X_{(j)}\bigr)\Bigr)}{\,n - j + 0.5} - \frac{\ln\Bigl(F\bigl(X_{(j)}\bigr)\Bigr)}{\,j - 0.5} \right\}^{\frac{n-0.5}{\,j-0.75} - \frac{1}{2}},
    \]
    where \(X_{(1)} \le X_{(2)} \le \cdots \le X_{(n)}\) denote the order statistics.
\end{enumerate}

\textbf{2. Meintanis Test Statistic (Meintanis, 2009):}
\[
\begin{split}
MEn = \frac{1}{n}\sum_{j=1}^{n}\sum_{k=1}^{n} \frac{2a}{(U_j - U_k)^2 + a^2} &+ 2\sum_{j=1}^{n} \Biggl[ \tan^{-1}\Bigl(\frac{U_j}{a}\Bigr) + \tan^{-1}\Bigl(\frac{1-U_j}{a}\Bigr) \Biggr] \\
&- 4n\,\tan^{-1}\Bigl(\frac{1}{a}\Bigr) - n\ln\Bigl(1+\frac{1}{a^2}\Bigr),
\end{split}
\]
where \(U_j = F(X_j)\) and the tuning parameter is set to \(a = 0.5\).

\textbf{3. Cramér--von Mises (CvM) Test Statistic:}
\[
CvM = \frac{1}{12n} + \sum_{j=1}^{n} \left( F\bigl(X_{(j)}\bigr) - \frac{2j-1}{2n} \right)^2.
\]

\textbf{4. Anderson--Darling (AD) Test Statistic:}
\[
AD = -n - \frac{1}{n}\sum_{i=1}^{n} \left[ (2i-1)\ln\Bigl( F\bigl(X_{(i)}\bigr) \Bigr) + (2n+1-2i)\ln\Bigl(1-F\bigl(X_{(i)}\bigr) \Bigr) \right].
\]

\textbf{5. Kolmogorov--Smirnov (KS) Test Statistic:}
\[
KS = \sup_{x \ge 1} \Bigl| F_n(x) - F(x) \Bigr|,
\]
where \(F(x) = 1 - x^{-\alpha}\) is the theoretical CDF under the null and \(F_n(x)\) is the empirical CDF.

\begin{table}[ht]
\centering
\caption{Alternative distributions and their density functions.}
\label{tab:alt_dists}
\begin{tabular}{@{}p{4.5cm}p{10.5cm}@{}}
\toprule
\textbf{Distribution} & \textbf{Density Function $f(x)$} \\
\midrule
Gamma &
$\displaystyle f(x) = \frac{1}{\Gamma(\lambda)} (x - 1)^{\lambda - 1} e^{-(x - 1)}, \quad x > 0, \quad \lambda >0  $\\
\addlinespace[0.75em]
Beta-Exponential &
$\displaystyle f(x) = \lambda\, e^{-(x - 1)} \left(1 - e^{-(x - 1)}\right)^{\lambda - 1}, \quad x > 0,\quad \lambda >0$ \\
\addlinespace[0.75em]
Tilted Pareto &
$\displaystyle f(x) = (1 + \lambda)(x + \lambda)^{-2}, \quad\quad x > 0,\quad \lambda >0$ \\
\addlinespace[0.75em]
Inverse Beta &
$\displaystyle f(x) = (1 + \lambda)(x - 1)^{\lambda}\, x^{-(2 + \lambda)}, \quad x > 0,\quad \lambda >0$ \\
\addlinespace[0.75em]
Benini &
$\displaystyle f(x) = x^{-2} \left(1 + 2\lambda \ln x\right) e^{-\lambda (\ln x)^2}, \quad x > 0,\quad \lambda >0$ \\
\addlinespace[0.75em]
Weibull &
$\displaystyle f(x) = \lambda\, x^{\lambda - 1} e^{-x^{\lambda}},\quad x > 0, \quad \lambda >0$ \\
\addlinespace[0.75em]
Half-Normal &
$\displaystyle f(x) = \sqrt{\frac{2}{\pi}}\, e^{-x^2/2},\quad x > 0$ \\
\addlinespace[0.75em]
Log-Normal &
$\displaystyle f(x) = \frac{1}{x \sqrt{2\pi}}\, e^{-\frac{(\ln x)^2}{2}},\quad x > 0$ \\
\bottomrule
\end{tabular}
\end{table}

\begin{table}[h!]
    \centering
    \caption{Comparison of empirical size and power for ($\alpha = 0.05$) }
    \begin{tabular}{c ccccccc}
        \toprule
        \hline
        $n$ & $\Delta$ & $KS$ & $CvM$ & $AD$ & $Z_{an}$ & $Z_{bn}$ & $M_{en}$ \\
        \hline
        \midrule
        \multicolumn{8}{l}{\textbf{Pareto(1)}} \\
        10 &0.044 & 0.065  & 0.070 &0.045 & 0.396  & 0.374  & 0.020 \\
        25 &0.045  & 0.048  &0.063   & 0.058 & 0.515&  0.029 & 0.006 \\
        50 &0.054  & 0.063  & 0.054  & 0.052  &  0.267&  0.466 & 0.004 \\
        75 &0.059  & 0.056  & 0.072  & 0.055  & 0.39 & 0.124 & 0.006 \\
        100 &0.049   &  0.047  & 0.050 & 0.024  & 0.329 & 0.605& 0.002 \\
        \midrule
        \multicolumn{8}{l}{\textbf{Gamma (0.5)}} \\
        10 & 0.899 & 0.603 & 0.669 & 0.498 & 0.354  & 0.344  & 0.291 \\
        20 & 0.990 & 0.827 & 0.907 & 0.734 & 0.383  & 0.373  & 0.275 \\
        30 & 1.000& 0.941 & 0.975 & 0.857& 0.446 & 0.474 & 0.326 \\
        40 & 1.000 & 0.956 & 0.992  & 0.916 & 0.515&  0.545& 0.413 \\
        50 & 1.000 & 0.985 & 0.998 & 0.953  & 0.554 &0.601 & 0.464 \\
        \midrule
        \multicolumn{8}{l}{\textbf{Beta Exponential(1)}} \\
        10 & 0.843 &0.651  &  0.633  & 0.740  & 0.493 & 0.544 & 0.166 \\
        20 & 0.976 & 0.848  & 0.888  & 0.914  & 0.571 & 0.498& 0.292 \\
        30 &0.993 & 0.939  & 0.965 & 0.977  & 0.666 & 0.558 & 0.409 \\
        40 & 1.000 & 0.974 & 0.988 & 0.993 & 0.737 & 0.539& 0.494 \\
        50 & 1.000 & 0.991 & 0.998 & 0.998 & 0.823 & 0.538 & 0.597 \\
        \bottomrule
    \end{tabular}
    \label{4h}
\end{table}
\begin{table}[h!]
    \centering
    \caption{Comparison of  power for ($\alpha = 0.05$) }
    \label{5h}
    \begin{tabular}{c ccccccc}
        \toprule
        \hline
        $n$ & $\Delta$ & $KS$ & $CvM$ & $AD$ & $Z_{an}$ & $Z_{bn}$ & $M_{en}$ \\
        \hline
        \midrule
        \multicolumn{8}{l}{\textbf{Inverse Beta(0.5)}} \\
        10 & 0.905 & 0.331 & 0.345  & 0.826 & 0.956 & 0.639& 0.018 \\
        20 & 0.965  & 0.395 & 0.452 & 0.968 & 0.996 & 0.687 & 0.084 \\
        30 & 0.976 & 0.415 & 0.543 & 0.991 & 1.000& 0.725& 0.121 \\
        40 & 0.981 & 0.472 & 0.689  & 0.998& 1.000 & 0.749 & 0.184 \\
        50 & 0.988 & 0.574  & 0.756 & 0.999 & 1.000 & 0.811& 0.239 \\
        \midrule
        \multicolumn{8}{l}{\textbf{Benini(1.5)}} \\
        10 & 0.767& 0.143 & 0.381 & 1.000 & 0.318 & 0.516 & 0.130 \\
        20 & 0.902 & 0.361 & 0.645 & 1.000& 0.277& 0.485& 0.250 \\
        30 & 0.971 & 0.611 & 0.784  & 1.000 & 0.325 & 0.550 & 0.335 \\
        40 & 1.000 & 0.752 & 0.887 & 1.000 & 0.333 & 0.535 & 0.429 \\
        50 & 1.000 & 0.833 & 0.942  & 1.000 & 0.401 & 0.530 & 0.549 \\
        \midrule
        \multicolumn{8}{l}{\textbf{Weibull(1.5)}} \\
        10 & 0.976 & 0.207 & 0.420 & 0.999 & 0.759 & 1.000 & 0.235 \\
        20 & 1.000 &  0.450 & 0.665 & 1.000 & 0.973 & 1.000 & 0.472 \\
        30 &  1.000 &  0.601 & 0.818 & 1.000 & 0.997 & 1.000 & 0.618 \\
        40 &  1.000 &  0.759 & 0.877 & 1.000 & 0.999 & 1.000 & 0.729 \\
        50 &  1.000 &  0.866 & 0.926 & 1.000 & 1.000 & 1.000 & 0.838 \\
        \midrule
        \multicolumn{8}{l}{\textbf{Half Normal (1) }} \\
        10 & 0.826 & 0.292 & 0.609 & 1.000 & 0.749 & 1.000 & 0.211 \\
        20 & 0.911 & 0.623 & 0.837 & 1.000 & 0.986 & 1.000 & 0.406 \\
        30 & 1.000 & 0.831 & 0.943 & 1.000 & 1.000 & 1.000 & 0.533 \\
        40 & 1.000 & 0.935 & 0.980 & 1.000 & 1.000 & 1.000 & 0.685 \\
        50 & 1.000 & 0.987 & 0.999 & 1.000 & 1.000 & 1.000 & 0.757 \\
        \midrule
        \multicolumn{8}{l}{\textbf{Log Normal(0.5)}} \\
        10 & 0.617 & 0.636 & 0.907 & 0.996 & 0.253 & 1.000 & 0.175 \\
        20 & 0.746 & 0.961 & 1.000 & 1.000 & 0.520 & 1.000 & 0.434\\
        30 & 0.817 & 0.999 & 1.000 & 1.000 & 0.704 & 1.000 & 0.633 \\
        40 & 1.000 & 1.000 & 1.000 & 1.000 & 0.852 & 1.000 & 0.770 \\
        50 & 1.000 & 1.000 & 1.000 & 1.000 & 0.912 & 1.000 & 0.861 \\
        \bottomrule
    \end{tabular}
\end{table}

Tables ~\ref{4h} and ~\ref{5h} provide information on the comparison results using a wide range of alternative distributions. It is observed that, as the sample size $n$ increases, the empirical size of the proposed test converges to the significance level. The analysis reveals that, in the majority of cases evaluated, the newly proposed tests exhibit superior performance compared to other tests.
\section{Data Analysis}

 The exceedances of flood maxima from the Wheat on River, located near
 Car cross in Canada’s Yukon Territory (Choulakian and Stephens (2001)),
 were examined. Table ~\ref{6h} shows the dataset used for this research, which
 contains 72 exceedance readings for the years 1958 and 1984. All values have
 been rounded to the nearest tenth of a cubic meter per second($m^3/s$).

\begin{table}[h]
    \centering
    \caption{Data set 1}
    \label{6h}
    \begin{tabular}{cccccccc}
        \toprule
        1.7 & 2.2  & 14.4 & 1.1  & 0.4  & 20.6 & 5.3  & 0.7  \\
        13.0 & 12.0 & 9.3  & 1.4  & 18.7 & 8.5  & 25.5 & 11.6 \\
        14.1 & 22.1 & 1.1  & 2.5  & 14.4 & 1.7  & 37.6 & 0.6  \\
        2.2  & 39.0 & 0.3  & 15.0 & 11.0 & 7.3  & 22.9 & 1.7  \\
        0.1  & 1.1  & 0.6  & 9.0  & 1.7  & 7.0  & 20.1 & 0.4  \\
        14.1 & 9.9  & 10.4 & 10.7 & 30.0 & 3.6  & 5.6  & 30.8 \\
        13.3 & 4.2  & 25.5 & 3.4  & 11.9 & 21.5 & 27.6 & 36.4 \\
        2.7  & 64.0 & 1.5  & 2.5  & 27.4 & 1.0  & 27.1 & 20.2 \\
        16.8 & 5.3  & 9.7  & 27.5 & 2.5  & 27.0 & 1.9  & 2.8  \\
        \bottomrule
    \end{tabular}
\end{table}

 Using the proposed goodness-of-fit test, we examine whether the given dataset follows a Pareto Type I distribution. The moment based estimator is $\hat{\alpha} = 1.0789$, and the observed value of the test statistic is $\hat{\Delta} = 0.2086$. At the 5\% significance level, the test fails to reject the null hypothesis $H_0$. Therefore, we conclude that the data follow a Pareto Type I distribution.
\\ \\

Next, we consider a dataset presented in Ahrari et al.\ (2022), which is assumed to follow the Rayleigh distribution with scale parameter 1. Table ~\ref{7h} shows the details of the dataset.

\begin{table}[h]
    \centering
    \caption{Data set 2 }
    \label{7h}
    \begin{tabular}{ccccc}
        \toprule
         0.2071766 & 0.6945765 & 1.0085693 & 1.0149304 & 1.1273867 \\ 1.2283711 & 1.3996847 & 1.4266420 & 1.8104736 & 1.8117200 \\ 1.8174535 & 1.9283017 & 2.1714312 & 2.5032170 & 2.7882901\\
        \bottomrule
    \end{tabular}
\end{table}
\FloatBarrier
For this dataset, the corresponding moment-based estimator is $\hat{\alpha} = 1.2557$. Using the proposed goodness-of-fit test, we examined and the value of the test statistic. The calculated test statistic $\hat{\Delta} = 0.6605$ indicates that at the significance level 5\%, the test leads to a rejection of the null hypothesis. Therefore, it is concluded that the dataset does not follow the Pareto type I distribution.
\doublespace
\section{Conclusion}
In this paper, we study the properties of the Weighted Entropy Generating Function (WEGF). A dynamic version of WEGF was introduced, Weighted Residual Entropy Generating Function (WREGF). Several properties and theorems related to WREGF were investigated, and characterization results were also explored. We have shown that WREGF uniquely determines the distribution. Further, we studied the relationships between WREGF, hazard rate, and mean residual life function. In addition, two new classes of lifetime distributions were introduced. A goodness-of-fit test for the Pareto distribution, based on the characterization result, was developed. The consistency of the proposed test was established using a linear combination of kernel density estimators. The power of the test was assessed through Monte Carlo simulations. The empirical type I of the test is well maintained. Furthermore, two real-life data sets were analyzed to evaluate whether the proposed test could accurately determine whether the data follow a Pareto distribution.

Several extensions of entropy and extropy measures have been investigated in the literature. The generating function framework offers further possibilities for exploring these measures. Moreover, empirical likelihood and jackknife empirical likelihood methods can be developed for their statistical inference.

\end{document}